\documentclass{arxiv}
\usepackage[T1]{fontenc}
\usepackage[latin9]{inputenc}
\usepackage{amstext}
\usepackage{graphicx}
\usepackage{multirow}

\title{Study of X-ray radiation damage in the AGIPD sensor for the European XFEL}

\author{Jiaguo Zhang$^{a,b}$\thanks{Corresponding
author.}~, Eckhart Fretwurst$^b$, Heinz Graafsma$^{a}$, Robert Klanner$^b$, Ioannis Kopsalis$^b$, and Joern Schwandt$^b$\\
\llap{$^a$}Deutsches Elektronen-Synchrotron (DESY), \\
  Notkestrasse 85, D-22607 Hamburg, Germany\\
\llap{$^b$}Institute for Experimental Physics, Hamburg University,\\
  Luruper Chaussee 149, D-22761 Hamburg, Germany\\
  E-mail: \email{jiaguo.zhang@desy.de}}

\abstract{The European X-ray Free Electron Laser (XFEL), currently being constructed in Hamburg and planning to be operational in 2017 for users, will deliver 27,000 fully coherent, high brilliance X-ray pulses per second with duration less than 100 fs. The unique features of the X-ray beam pose major challenges for detectors used at the European XFEL for imaging experiments, in particular a radiation tolerance of silicon sensors for doses up to 1 GGy for 3 years of operation at an operating voltage above 500 V.\\
\\
One of the detectors under development at the European XFEL is the Adaptive Gain Integrating Pixel Detector (AGIPD), which is a hybrid detector system with ASICs bump-bonded to p$^{+}$n silicon pixel sensors. We have designed the silicon sensors for the AGIPD, which have been fabricated by SINTEF and delivered in the beginning of February of 2013. To demonstrate the performance of the AGIPD sensor with regard to radiation hardness, mini-sensors with the same pixel and guard-ring designs as the AGIPD together with test structures have been irradiated at the beamline P11 of PETRA III with 8 keV and 12 keV monoenergetic X-rays to dose values up to 10 MGy. The radiation hardness of the AGIPD sensor has been proven and all electrical properties are within specification before and after irradiation. In addition, the oxide-charge density and surface-current density from test structures have been characterized as function of the X-ray dose and compared to previous measurements for test structures produced by four vendors.}

\keywords{XFEL; silicon pixel sensor; X-ray radiation damage; radiation hardness}

\begin{document}


\section{Introduction}

\begin{figure}[htbp]
\small
\centering
\includegraphics[scale=0.5]{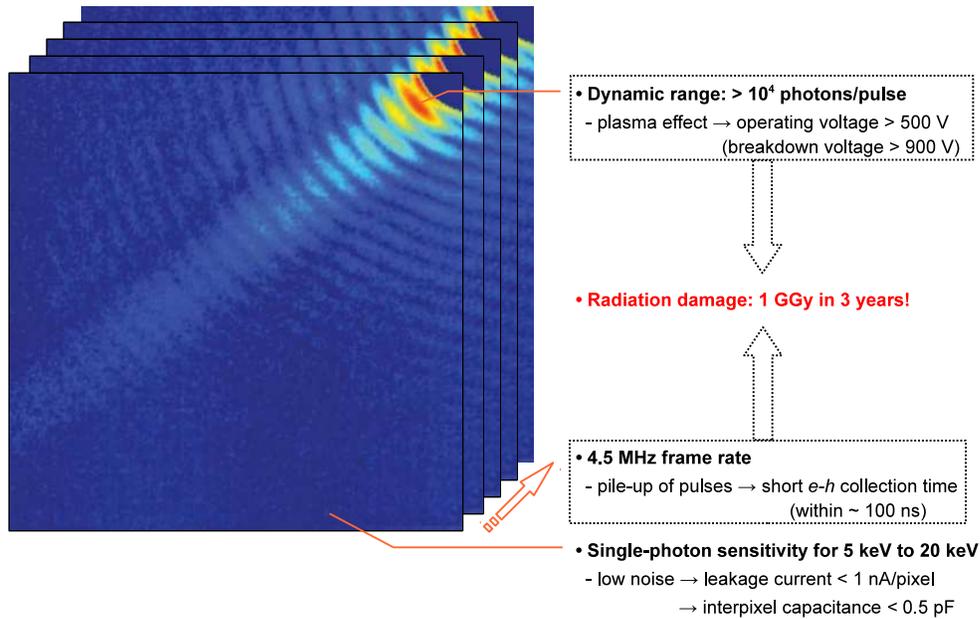}
\caption{Challenges of silicon detectors used for imaging experiments at the European XFEL. Picture reproduced from \cite{bib2}.}
\label{Challenges}
\end{figure}

The European X-ray Free-Electron Laser (XFEL) \cite{bib1}, currently being constructed in Hamburg and planned to be operational for users in 2017, will deliver 27,000 fully coherent, high brilliance X-ray pulses per second with duration less than 100 fs and time separation of 220 ns. The unique features of the X-ray beam pose major challenges for detectors used at the European XFEL for imaging experiments, as shown in figure \ref{Challenges}: A dynamic range of 0, 1, ..., up to more than 10$^{4}$ photons per pulse, a frame rate of 4.5 MHz, and in particular a radiation tolerance of the sensors for doses up to 1 GGy for 3 years of operation.

One of the detectors under development at the European XFEL is the Adaptive Gain Integrating Pixel Detector (AGIPD) \cite{bib3, bib4}. It is a hybrid detector system with 1 Mega pixels, each of \mbox{200 $\times$ 200 $\mu$m$^{2}$}. The AGIPD detector consists of 16 modules, each with 16 ASICs bump-bonded to one p$^{+}$n silicon pixel sensor. The AGIPD sensor has been designed according to the optimized pixel and guard-ring geometries and technological processes based on \textsc{SYNOPSYS} TCAD simulations \cite{bib5,bib6,bib7} with damage-related parameters \cite{bib8, bib9} implemented. The sensors have been produced by SINTEF \cite{bib10} and delivered at the beginning of February 2013. The basic properties of the silicon wafers have been obtained from electrical measurements on test structures. 

In order to demonstrate the radiation hardness of the AGIPD sensor, mini-sensors with the same pixel and guard-ring designs as the AGIPD and test structures have been irradiated up to \mbox{10 MGy} at the beamline P11 of PETRA III at DESY. From the test structures, the oxide-charge density, $N_{ox}$, and surface-current density, $J_{surf}$, have been extracted as a function of dose. The results have been compared to previous ones: A saturation of $N_{ox}$ and $J_{surf}$ has been found between 100 kGy and 1 MGy and the saturation values are smaller than those from test structures produced by other vendors. From mini-sensors, no breakdown has been found up to 900 V after irradiation and all electrical properties, in particular the leakage current and inter-pixel capacitance, are well within specifications. However, it is found that the mini-sensors after wafer cutting show an early soft breakdown at about 820 V, which finally are recovered after irradiation to 200 Gy.

In the following, the requirements of the AGIPD sensor for experiments at the European XFEL and the design of the AGIPD sensor will be shortly adressed and results of the test structures and the AGIPD mini-sensors from radiation tests discussed.

\section{Requirements and design of the AGIPD sensor}

The AGIPD requires a silicon sensor with a thickness of 500 $\mu$m in order to obtain a high quantum efficiency for X-rays with energies between 5 keV and 20 keV \cite{bib11}. At the European XFEL, the high intensity of interacting photons produces an electron-hole plasma in the 500 $\mu$m thick silicon sensor after a single shot. In order to avoid pile-up effects the generated electrons and holes have to be collected within $\sim$ 100 ns and, therefore, an operating voltage of \mbox{> 500 V} is needed \cite{bib12, bib13, bib14}. Thus, a breakdown voltage of the AGIPD sensor of \mbox{> 900 V}, which is above the operating voltage, is specified. In addtion, to limit the noise in order to achieve a single photon sensitivity for X-rays down to an energy of 12.4 keV, a single pixel leakage current of < 1 nA and an interpixel capacitance of < 0.5 pF, is required. The aforementioned specifications have to be satisfied for a period of three years, during which the sensor will be exposed to a dose in the SiO$_{2}$ of up to 1 GGy. Thus, the AGIPD sensor has to be optimized taking into account the effects of radiation damage.

X-rays with energies below 300 keV will only produce surface damage in the SiO$_{2}$ insulating layer and at the Si-SiO$_{2}$ interface of the silicon sensor: Oxide charges and interface traps will build up with X-ray irradiation \cite{bib15, bib16}. In general, the former induces electron-accumulation layers below the Si-SiO$_{2}$ interface in-between neighbouring p$^{+}$-implant electrodes in p$^{+}$n sensor and causes an early breakdown of the sensor; the latter contributes to the surface current. The oxide-charge density and surface-current density have been previously characterized as function of dose up to 1 GGy \cite{bib9, bib17, bib18}. These parameters have been implemented into SYNOPSYS TCAD simulations for the optimization of different pixel geometries, guard-ring geometries and technological process parameters. It has been found that, in order to satisfy the requirements for pixel leakage current and interpixel capacitance after irradiation to 1 GGy dose, a gap of 20 $\mu$m between neighbouring p$^{+}$-implant electrodes and an aluminium metal overhang of 5 $\mu$m are needed. In order to achieve a breakdown voltage of \mbox{> 900 V}, an optimized p$^{+}$-implant depth of 2.4 $\mu$m, an oxide thickness of 250 nm, and at least of 15 guard rings are necessary. These 15 guard rings have to be arranged in a way that the voltage drops from one guard ring to another are equalised so that same electric fields at each guard-ring implant is achieved. The width of the inactive region at the edges of the AGIPD is 1.2 mm, where the 15 guard rings are implemented. The AGIPD sensor is covered by a passivation layer consisting of 500 nm SiO$_{2}$ and 250 nm Si$_{3}$N$_{4}$. Details on the AGIPD pixel and guard-ring designs can be found elsewhere \cite{bib6, bib7}.

\section{Investigated structures and properties of the AGIPD wafer}

A square pad diode with an active area of 5.05 mm $\times$ 5.05 mm on the AGIPD wafer has been used to investigate the full depletion voltage and the leakage current from bulk silicon for the AGIPD sensor. From capacitance-voltage (C-V) characteristic of the pad diode, a full depletion voltage of 95 V is obtained, which corresponds to a doping concentration of $5.3 \times 10^{11}$ cm$^{-3}$ or a resistivity of \mbox{$\sim$ 7.8 k$\Omega\cdot$cm}, respectively. The current-voltage (I-V) measurement shows a leakage current of \mbox{2.4 nA/cm$^{3}$} at 500 V. The low current density indicates a lifetime of the order of a few hundred miliseconds for electrons and holes in the silicon bulk. Such long lifetimes lead to the conclusion that the current is dominated by the diffusion current. 

\begin{figure}[htbp]
\small
\centering
\includegraphics[scale=0.52]{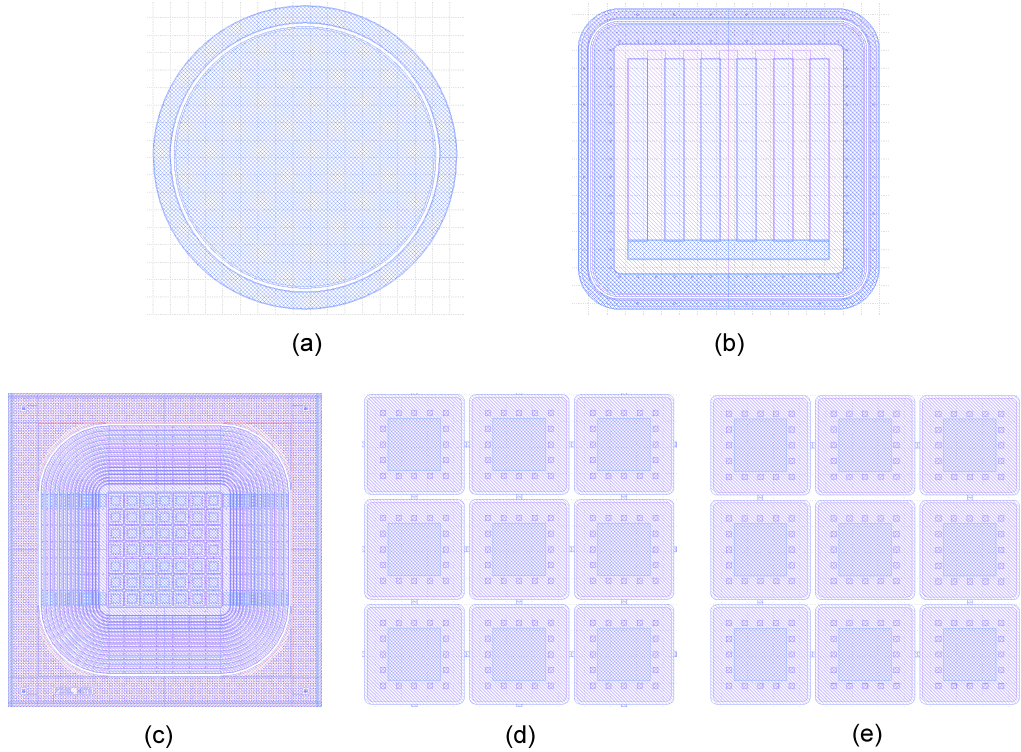}
\caption{Masks of (a) MOS capacitor, (b) gate-controlled diode, (c) entire mini-sensor, (d) central pixels of "inter-connected" mini-sensor (all pixels connected with aluminium bridges), (e) central pixels of "ring-connected" mini-sensor (pixels surrounding central pixel connected with aluminium bridges).}
\label{Masks}
\end{figure}

In addition, test fields, each including a metal-oxide-semiconductor (MOS) capacitor and a gate-controlled diode (GCD), have been designed and fabricated on the AGIPD wafer for the investigation of the damage-related parameters like the oxide-charge density and surface-current density. The MOS capacitor, shown in figure \ref{Masks}(a), has a circular shape with a diameter of 1.5 mm, surrounded by a 100 $\mu$m wide gate ring which was floating during the measurement. The C-V measurement on the MOS capacitor shows a flatband voltage ($V_{fb}$) of $-0.40$ V, which is smaller than the working function difference ($\phi_{ms}$) between aluminium and silicon of $-0.42$ V. The difference between $V_{fb}$ and $\phi_{ms}$ indicates a negtive charge in the insulating layer with an equivalent density of $1.8 \times 10^{10}$ cm$^{-2}$ at the Si-SiO$_{2}$ interface. The geometry of the gate-controlled diode is shown in figure \ref{Masks}(b): It is a finger-like structure with 6 vertical and 1 horizontal fingers of gates surrounded by a diode. The width of the 7 fingers is \mbox{100 $\mu$m} each and the lengths for the 6 vertical and 1 horizontal figures are 1000 $\mu$m and 1100 $\mu$m, respectively. This gives a gate area of $7.1\times  10^{-3}$ cm$^{2}$. The surface-current density from an I-V measurement on the gate-controlled diode is 2.0 nA/cm$^{2}$ for the AGIPD wafer before irradiation.

\begin{figure}[htbp]
\small
\centering
\includegraphics[width=7.5 cm]{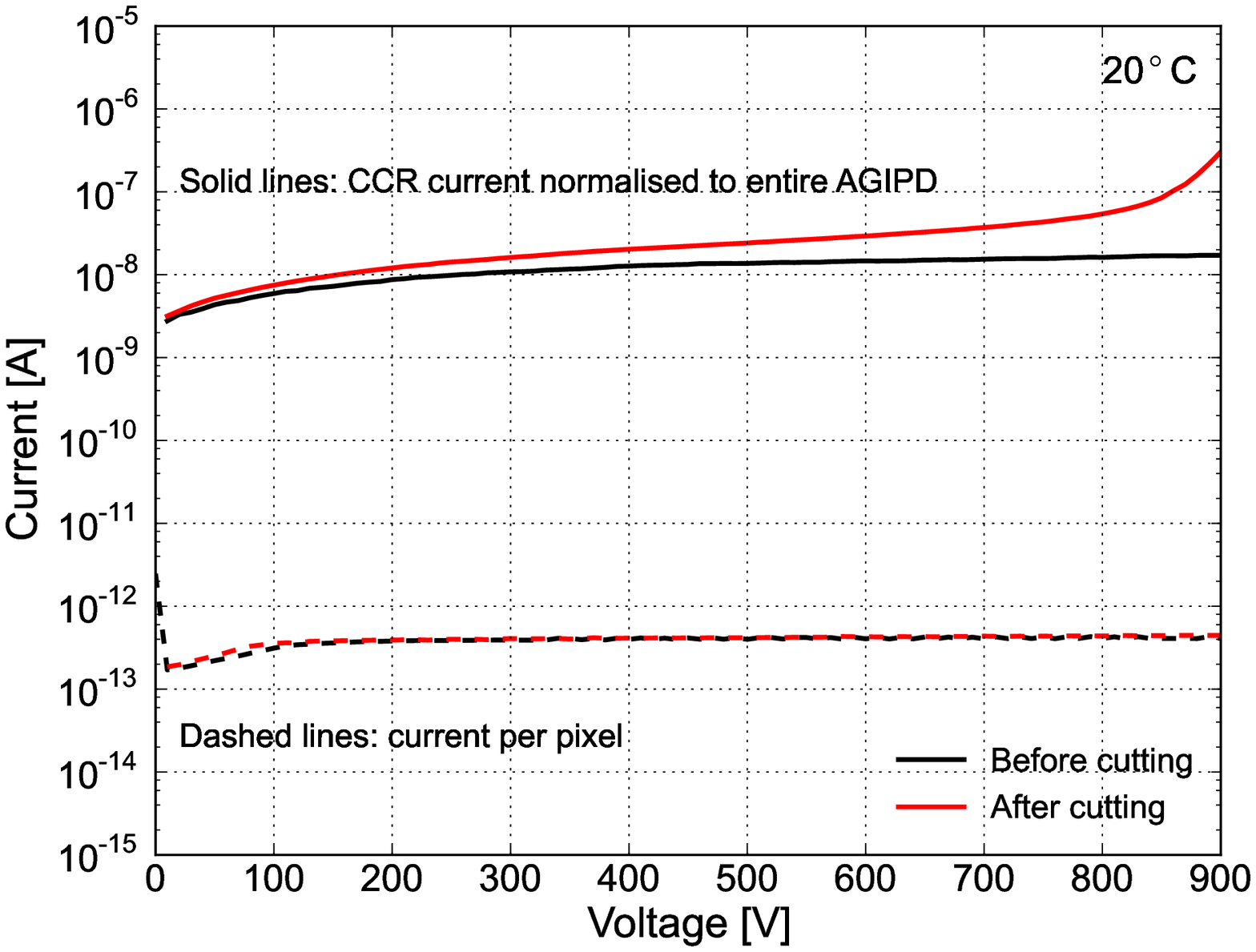}
\includegraphics[width=7.5 cm]{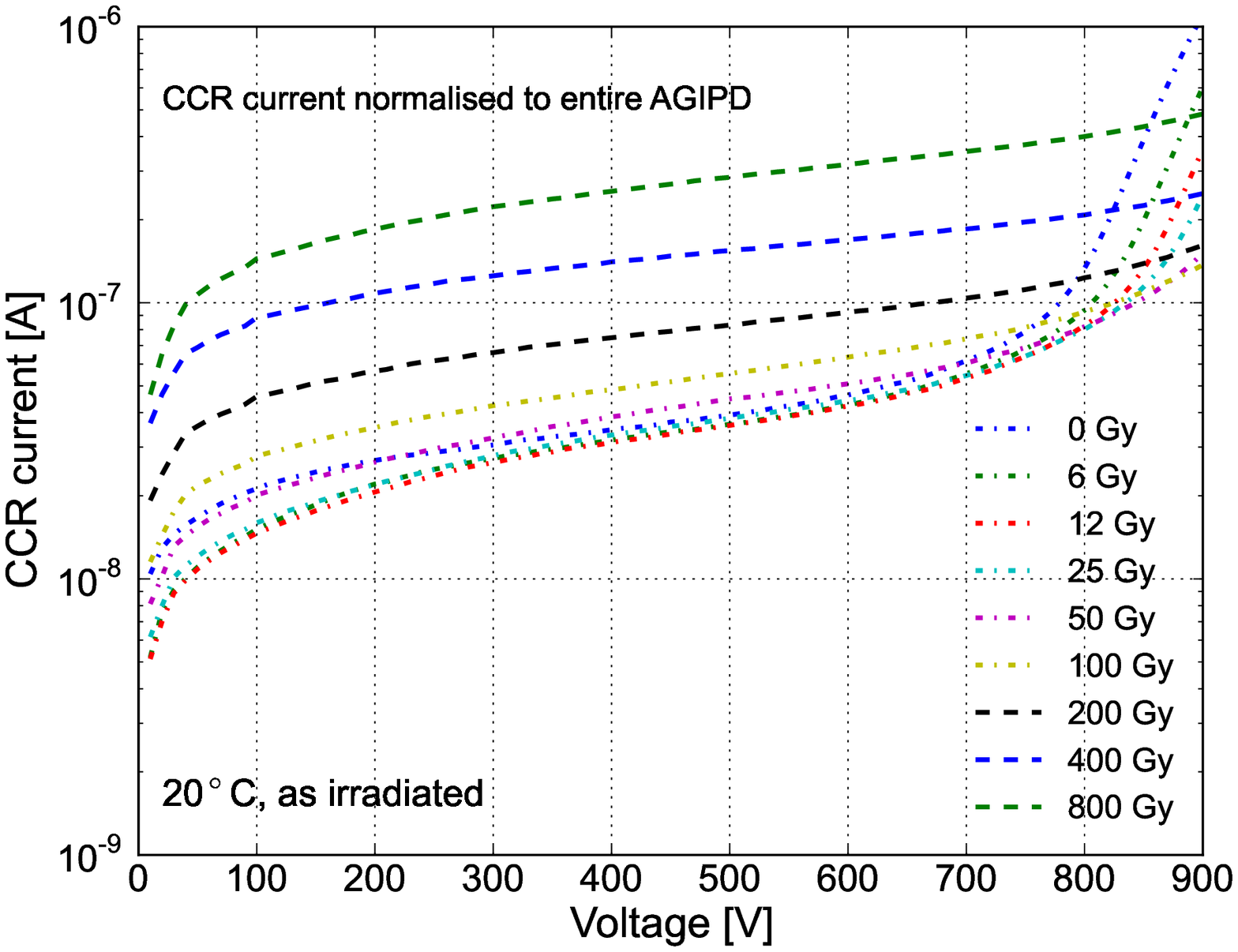}
\caption{Left: Dashed lines and solid lines represent a single pixel current and CCR current scaled to the entire AGIPD sensor before and after cutting. After cutting, a soft breakdown of $\sim$ 820 V is observed. Right: I-V curves of the CCR as function of X-ray dose. It is seen that above a dose of 200 Gy the soft breakdown at $\sim$ 820 V disappears.}
\label{IVs_cutting}
\end{figure}

To investigate the radiation hardness of the AGIPD sensor, two types of mini-sensors with the same pixel and guard-ring geometries as the AGIPD sensor but with only 7 $\times$ 7 pixels have been designed and fabricated on the same wafer, as shown in figure \ref{Masks}(c). The differences for the two types of mini-sensors are: One has aluminium bridges connecting all pixels together, called ``inter-connected'' mini-sensor (figure \ref{Masks}(d)); the other has aluminium bridges connecting those pixels in a ring shape surrounding the central pixel, called ``ring-connected'' mini-sensor (figure \ref{Masks}(e)). Such designs are used for the pixel leakage current and interpixel capacitance measurements. For the non-irradiated mini-sensors, the pixel current and the Current Collection Ring (CCR) current after scaling to the entire AGIPD sensor are \mbox{0.42 pA/pixel} and \mbox{21 nA}, respectively. There is no breakdown observed for the mini-sensors before cutting. However, a soft breakdown of the CCR at $\sim$ 820 V has been observed after cutting.  Figure \ref{IVs_cutting} (left) shows the I-V curves from a single pixel and the CCR before and after cutting. The early breakdown of CCR at $\sim$ 820 V is due to the depletion region reaching the cut-edge, where the crystal is damaged. This kind of breakdown can be healed through a low-dose irradiation with an in-house X-ray source. The irradiation increases the positive charges in the SiO$_{2}$ and prevents that the depletion region reaches the cut edge: Without irradiation, the depletion region reaches the n$^{+}$-implant at the Si-SiO$_{2}$ interface, which surrounds the entire sensor, due to the negative charge in the SiO$_{2}$; With increasing positive oxide-charge density, the depletion region ends at the interface region between the last guard ring and the n$^{+}$-implant and thus prevents further depletion in the silicon bulk to the cut-edge. Figure \ref{IVs_cutting} (right) shows the I-V curves of the CCR as function of the X-ray dose. Above 200 Gy, shown as black dashed line, there is no breakdown of CCR for voltages below 900 V.

For the investigation of the dose dependence of damage-related parameters, like the oxide-charge density and surface-current density, and of electrical properties of the AGIPD sensor, test structures and mini-sensors have been irradiated to 10 kGy, 100 kGy, 1 MGy and 10 MGy. Especally, 6 mini-sensors (3 for each type) have been irradiated in 3 different ways: One irradiated uniformly over the full sensor, one irradiated uniformly over half of the sensor, the other irradiated uniformly over the full sensor with 10 V bias voltage applied during irradiation.

\section{High-dose irradiations at PETRA III}

For the irradiations with X-rays, the set-up discussed in \cite{bib18, bib19} has been used at the beamline P11 of PETRA III. The set-up consists of an adjustable Ta collimator, which is used to precisely define the region of irradiation on the sensor, and a sample holder, which is connected to a liquid cooling system. The mini-sensors and test fields planned to be irradiated were glued onto a special designed ceramic and wire-bonded to the 5 biasing lines of the substrate. The ceramic substrates were mounted into the sample holder during irradiation.

\begin{figure}[htbp]
\small
\centering
\includegraphics[width=7.0 cm]{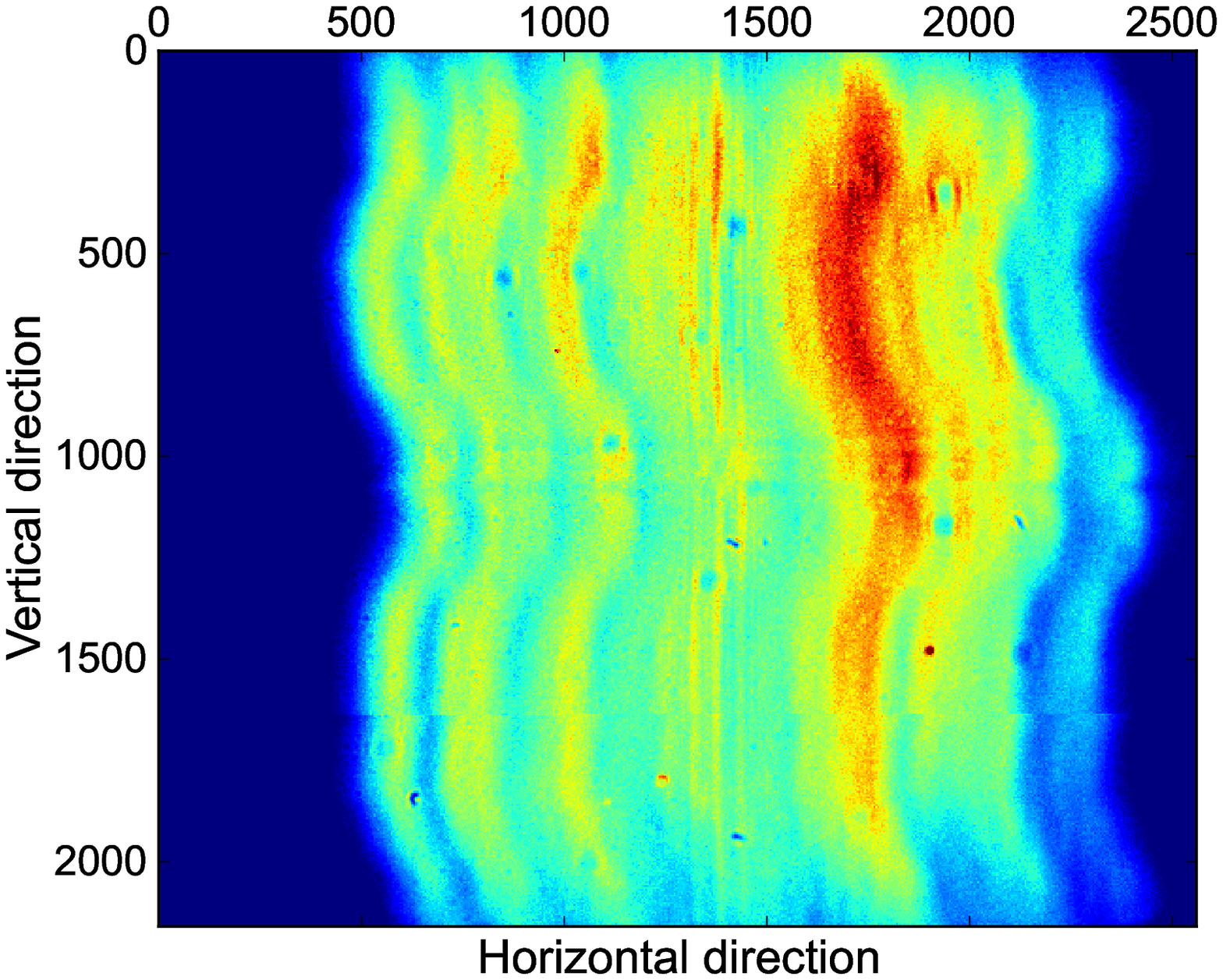}
\includegraphics[width=7.5 cm]{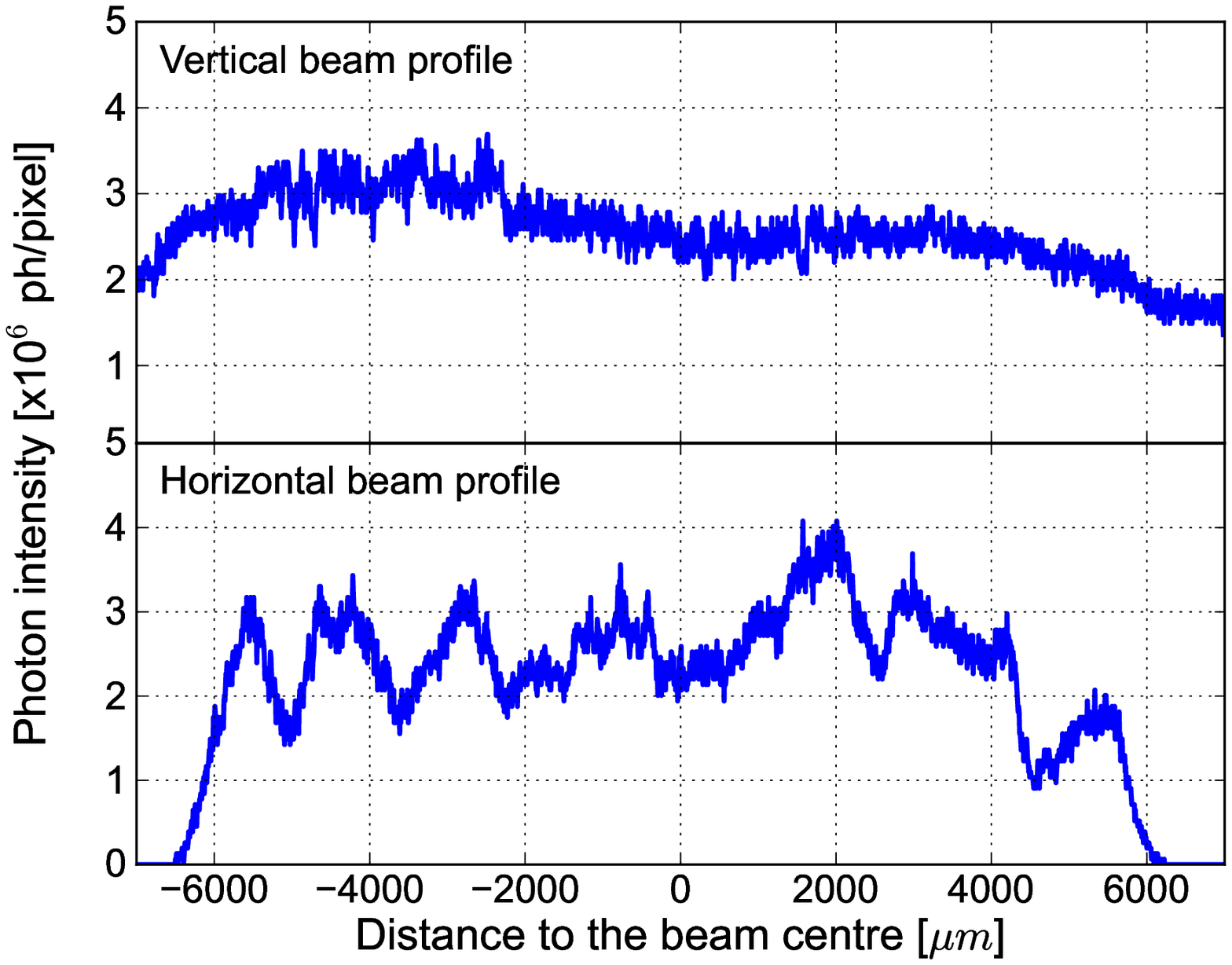}
\caption{Left: Image of the direct beam at the beamline P11 of PETRA III. Right: Horizonal and vertical beam profiles along the cuts through the beam centre.}
\label{BeamProfile}
\end{figure}

The beamline P11 at PETRA III provides monoenergetic X-ray beam from an undulator. The beam profile has been taken by using a CMOS camera (pco.edge 5.5 \cite{bib20}) with a resolution of 2560 $\times$ 2160 pixels, each of \mbox{6.5 $\mu$m} pitch, and a readout speed of 100 frames per second. \mbox{Figure \ref{BeamProfile}} shows the image from the direct beam taken with the camera and the profiles along the horizontal and vertical cuts through the beam centre. The size of the beam spot from the image is approximately 1.2 mm $\times$ 1.4 mm.

The photon intensity was calibrated with a silicon photodiode. The area of the photodiode is 1.0 cm $\times$ 1.0 cm, which is large enough to cover the entire beam. The photo-current from the \mbox{300 $\mu$m} thick photodiode for 8 keV X-rays was 3.5 mA, which corresponds to a photon intensity of $1\times 10^{13}$ photons/s obtained from a direct conversion: 

\begin{equation}
I_{x-ray}=\frac{3.6\textrm{eV} \cdot I_{diode}}{q_{0} \cdot E_{x-ray} \cdot (1-\textrm{exp}^{-\frac{T_{si}}{L_{x-ray}}})}
\end{equation}

\noindent where $I_{diode}$ is the photo-current, $q_{0}$ the elementary charge, $E_{x-ray}$ the energy of X-rays, $T_{si}$ the thickness of the photodiode and $L_{x-ray}$ the X-ray attenuation length for $E_{x-ray}$. Thus, the average dose rate in SiO$_{2}$ within the beam spot was 26 kGy/s for 8 keV X-rays. As the size of the mini-sensors and test fields are larger than the beam spot, irradiations by scanning the entire beam were needed in order to obtain an uniform irradiation. In this work, a ``square waveform''-shape irradiation path has been taken: The carrier stage of the sample holder was moved along the horizontal direction with a constant speed $V_{x}$ from one end to the other; the stage was then moved vertically by a small step of $Z_{step}$ at the end and then moves back horizontally. The accumulated dose is proportional to the dose rate integrated over entire beam area, and divided by the moving speed $V_{x}$ and the vertical step $Z_{step}$. Details of dose calculation are documented in \cite{bib18}. In this work, a constant speed $V_{x}$ of 1.0 mm/s during scans were used; however, the vertical step of $Z_{step}$ was changed to obtain different doses. During irradiation, the cooling temperature was set to 15$^{\circ}$C.

The mini-sensors and test fields were irradiated to accumulated doses of 10 kGy, 100 kGy, \mbox{1 MGy} and 10 MGy\footnote{12 keV X-rays were used for irradiations to 10 kGy, 100 kGy and 1 MGy, 8 keV X-rays used for 10 MGy.}. After each irradiation dose, their electrical properties were measured.

\section{Results}

\subsection{Dose dependence of damage-related parameters from test fields}

We first present the results from the MOS capacitor and gate-controlled diode irradiated to four \mbox{X-ray} doses. During the irradiations, no voltages were applied to the electrodes of the MOS capacitor and the gate-controlled diode. The first C/G-V and I-V measurements were performed within 1 hour after each irradiation. Measurements were also done after annealing at 80$^{\circ}$C for 10 minutes in order to obtain reproducible results \cite{bib8, bib18, bib21}.

\begin{figure}[htbp]
\small
\centering
\includegraphics[width=7.5 cm]{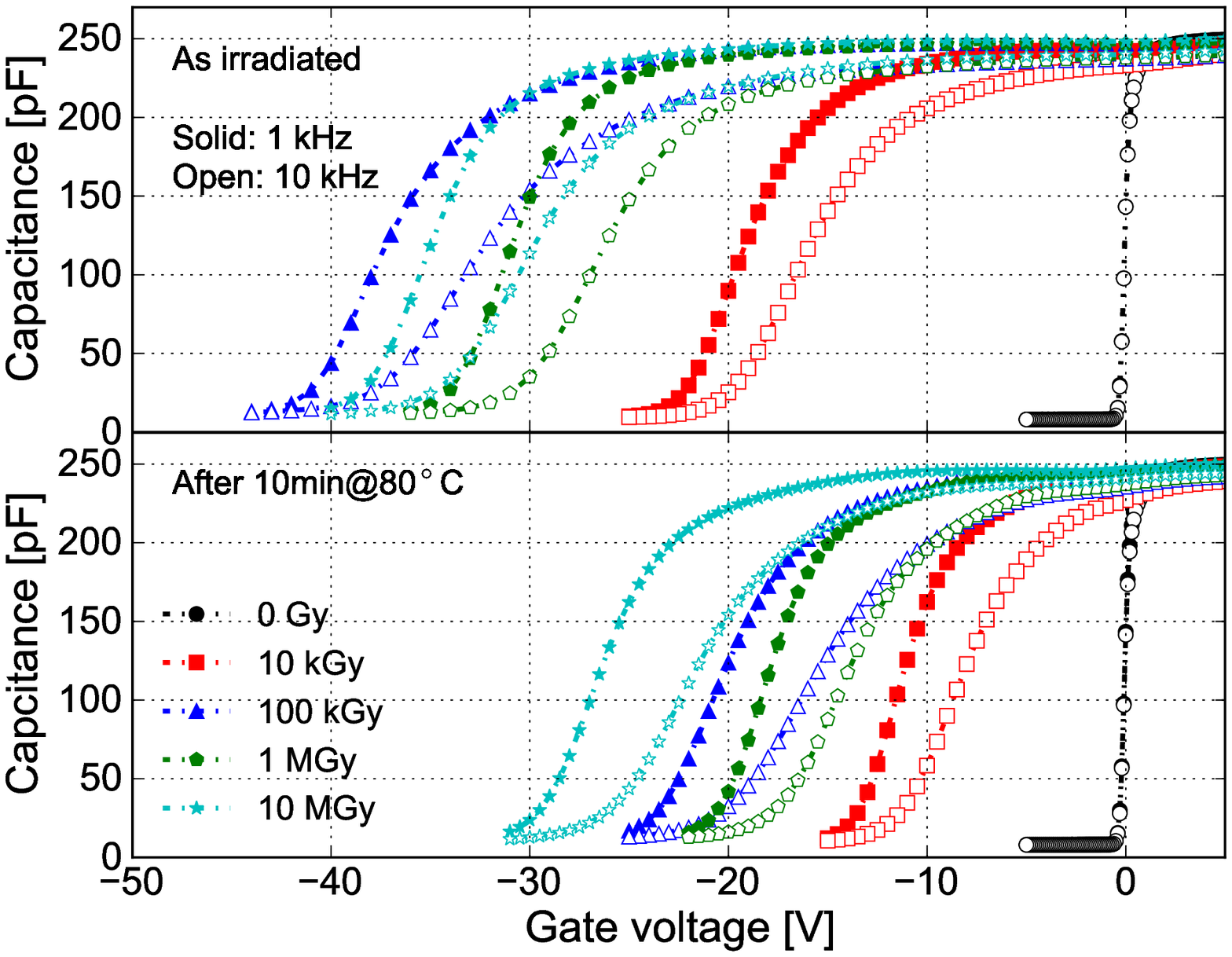}
\includegraphics[width=7.5 cm]{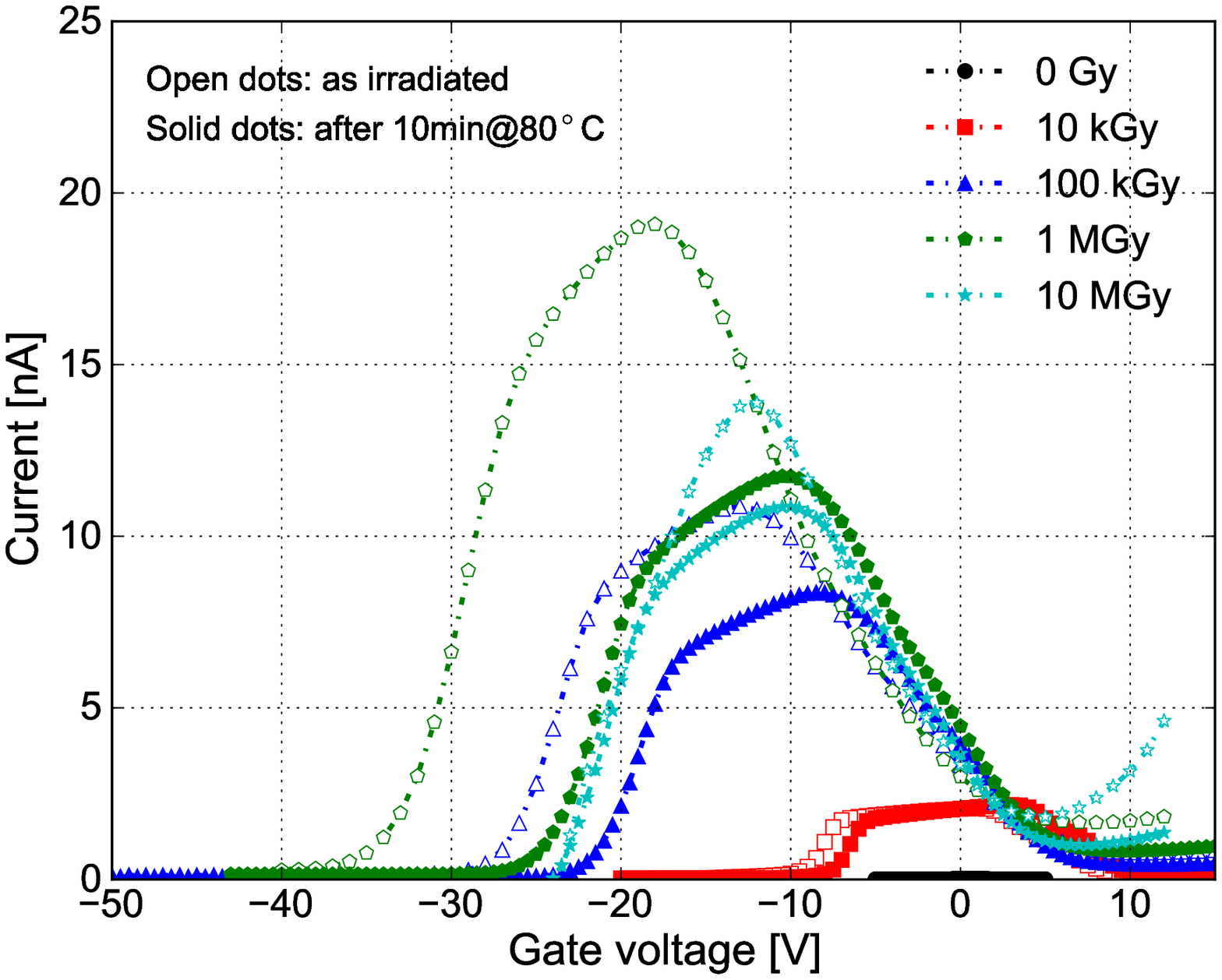}
\includegraphics[width=7.5 cm]{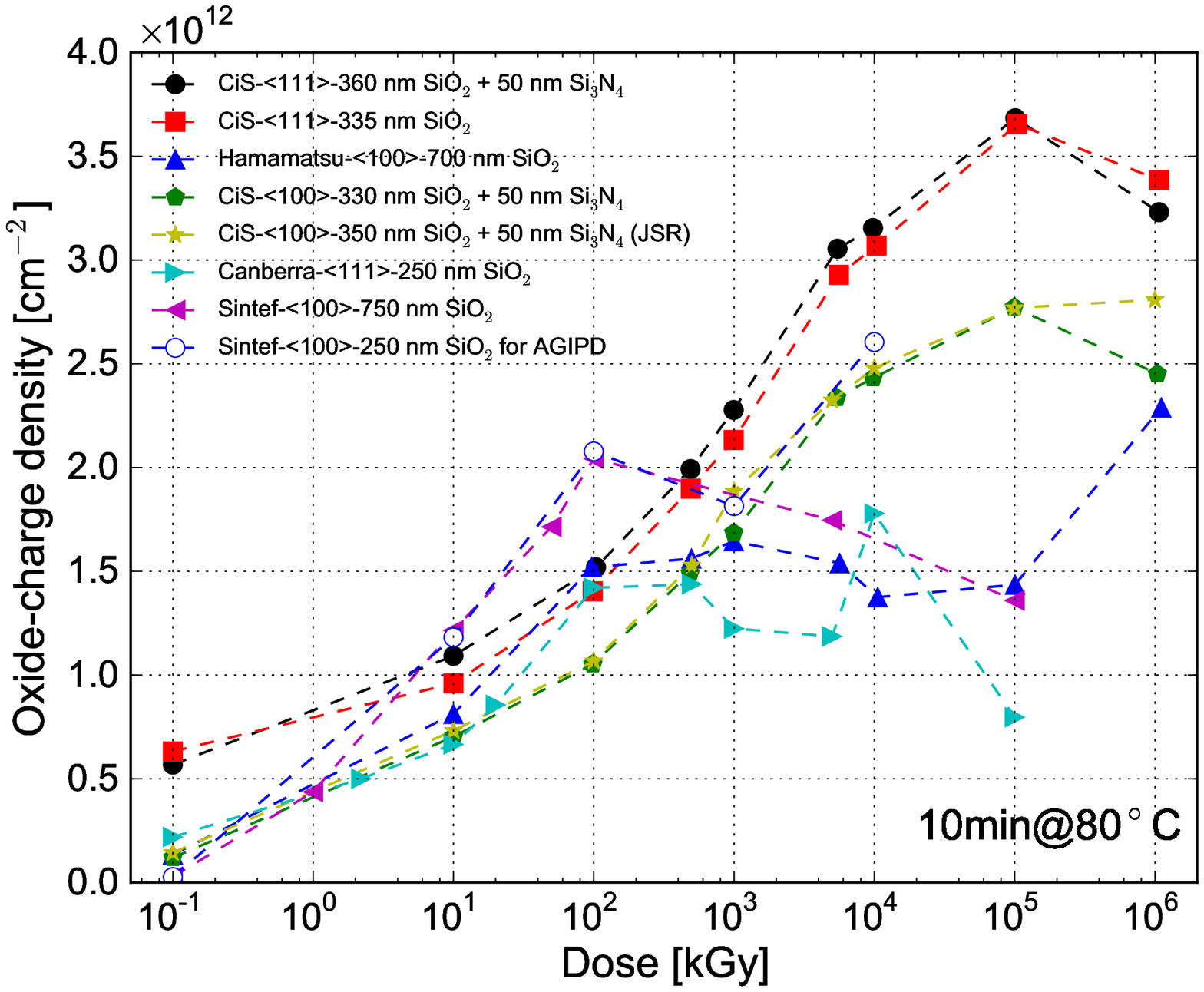}
\includegraphics[width=7.5 cm]{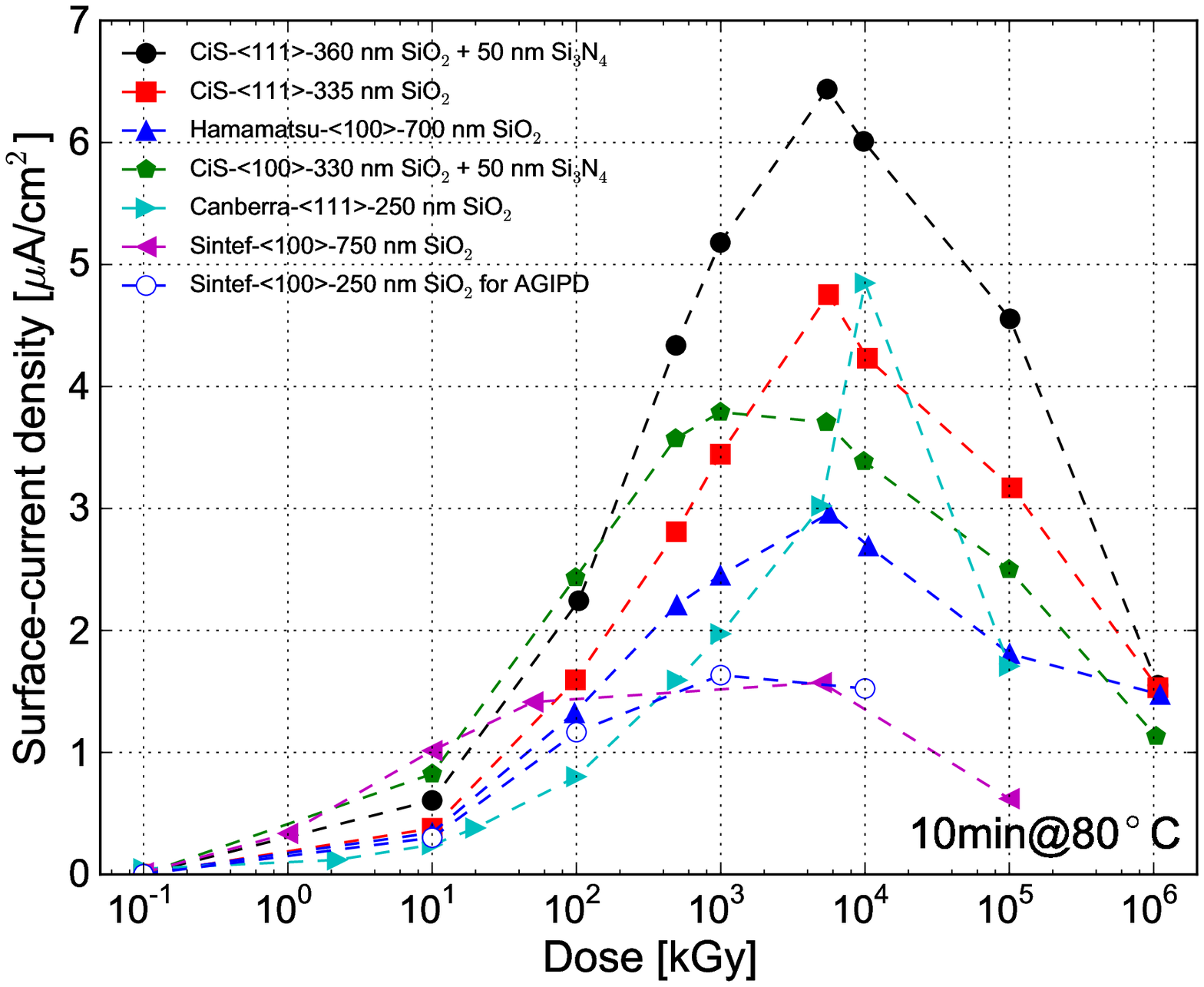}
\caption{Results from the MOS capacitor and gate-controlled diode. Top left: C-V curves of the MOS capacitor; Top right: I-V curves of the gate-controlled diode; Bottom left: Oxide-charge density vs. dose and comparison to previous measurements \cite{bib9, bib17, bib18}; Bottom right: Surface-current density vs. dose and comparison to previous results \cite{bib9, bib17, bib18}.}
\label{TSresults}
\end{figure}

On top of figure \ref{TSresults}, C/G-V and I-V curves from the MOS capacitor and gate-controlled diode before and after annealing at 80$^{\circ}$C for 10 minutes are shown. From the C/G-V curves of the as-irradiated MOS capacitor, a maxmimal voltage shift of 45 V has been found for 100 kGy. However, after annealing at 80$^{\circ}$C for 10 minutes, C/G-V curves from 10 MGy show a maximal voltage shift, which may indicate different annealing time constants for different doses: not only the annealing of oxide-trapped charges (conventionally what we called oxide charges), but also the removal of free carriers in the SiO$_{2}$ produced by prompt X-rays, which also can shift the flatband voltage of the MOS capacitor. From the I-V curves of the gate-controlled diode, the maximal surface current was found at 1 MGy and it decreases for doses above 1 MGy. On the bottom of figure \ref{TSresults}, oxide-charge density and surface-current density, extracted from the MOS capacitor and gate-controlled diode after annealing at 80$^{\circ}$C for 10 minutes and shown as blue open dots, are plotted as function of dose and compared to previous measurement results \cite{bib9, bib17, bib18}. It is found that $N_{ox}$ and $J_{surf}$ for the AGIPD wafer are consistent with the results obtained from the structures produced by the same vendor (SINTEF) with a different oxide thickness, shown in magenta rhombus. This confirms a previous conclusion that X-ray damage-related parameters, especially the oxide-charge density, weakly depend on the thickness of the SiO$_{2}$ \cite{bib9, bib18}, as most of the oxide-trapped charges are located within a few nanometers close to the Si-SiO$_{2}$ interface. $N_{ox}$ and $J_{surf}$ saturate at a dose between 100 kGy and 1 MGy, and the saturation values are $(2-2.5) \times 10^{12}$ cm$^{-2}$ and \mbox{1.7 $\mu$A/cm$^{2}$}.

\subsection{Dose dependence of electrical properties of the AGIPD mini-sensors}

Three pairs of ``inter-connected'' and ``ring-connected'' mini-sensors have been irradiated in different ways, as introduced in a previous section: One irradiated uniformly over the full sensor, one irradiated uniformly over half of the sensor, the other irradiated uniformly over the full sensor with 10 V bias voltage. The ``inter-connected'' mini-sensors are used for the investigation of the pixel current, CCR current and breakdown voltage, whereas the ``ring-connected'' mini-sensors for the interpixel capacitance.

\subsubsection{Pixel current, CCR current and breakdown voltage}

The I-V characteristics of each inter-connected mini-sensor were measured for each dose at 20$^{\circ}$C within one hour after irradiation and after annealing for 10 minutes at 80 $^{\circ}$C. In addition, measurements at $-$20$^{\circ}$C have also been done after annealing.

\begin{figure}[htbp]
\small
\centering
\includegraphics[width=7.5 cm]{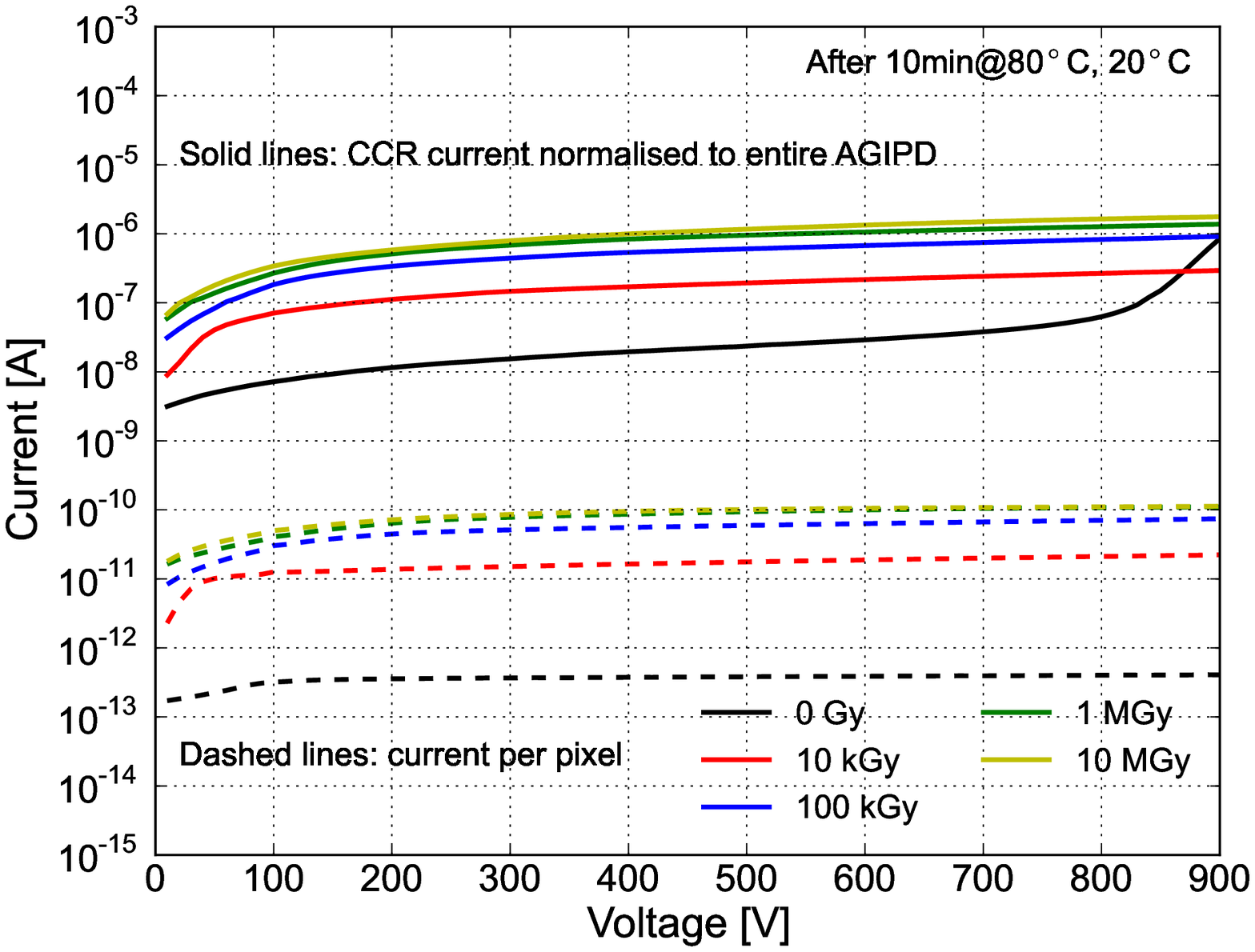}
\includegraphics[width=7.5 cm]{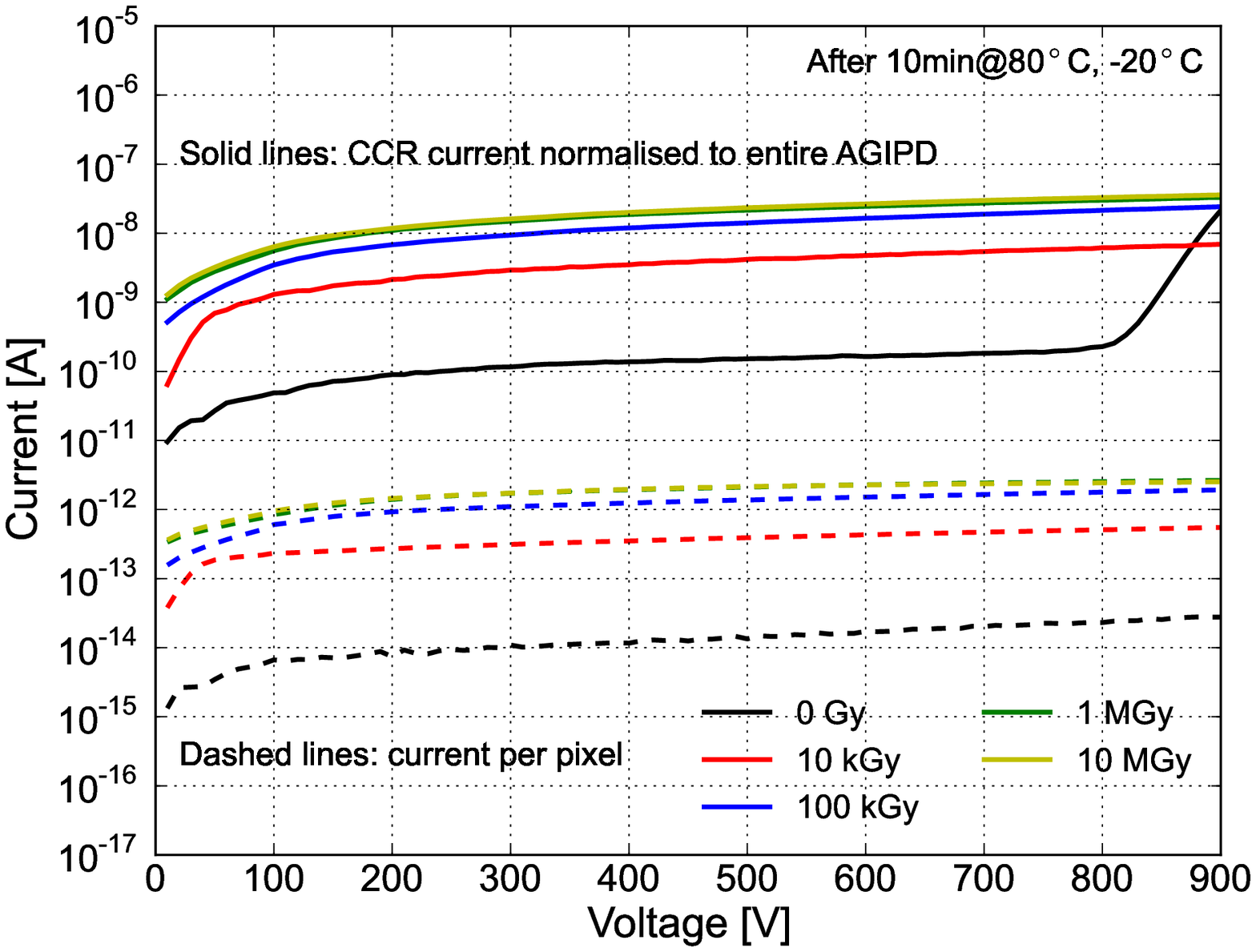}
\caption{Pixel current and CCR current as function of dose. Left: Measurements at 20$^{\circ}$C; Right: Measurements at $-$20$^{\circ}$C. CCR currents have been normalised to the entire sensor.}
\label{IVs_sensor}
\end{figure}

Figure \ref{IVs_sensor} shows the pixel current and CCR current from the ``inter-connected'' mini-sensors with uniform irradiation over the entire sensor measured at 20$^{\circ}$C and $-$20$^{\circ}$C as function of dose. Measurements have been done in a dry atmosphere with a relative humidity (RH) less than 3\%. Results have been scaled to the values for a single pixel and the entire CCR. Before annealing, not shown here, both pixel and CCR currents measured at 20$^{\circ}$C are less than 10\% higher than the values obtained after annealing. No breakdown has been observed. After annealing, as seen in \mbox{figure \ref{IVs_sensor}}, there is still no breakdown observed in the temperature range between $-$20$^{\circ}$C and 20$^{\circ}$C. The pixel and CCR currents are reduced by two orders of magnitude at $-$20$^{\circ}$C compared to at 20$^{\circ}$C. All current values up to 10 MGy satisfy the requirements for the AGIPD sensor: pixel current \mbox{< 1 nA} and CCR current \mbox{< 20 $\mu$A}. As both saturate at $\sim$ 1 MGy, no further changes are expected for doses above 10 MGy.

As expected, pixel and CCR currents of the ``inter-connected'' mini-sensors irradiated uniformly over half of the sensor are about 50\% of the values shown in figure \ref{IVs_sensor}. For such kind of irradiation, which simulates the more realistic case for the sensor at the European XFEL, no breakdown has been observed after each dose. In addition, the difference between the results obtained from the ``inter-connected'' mini-sensors with 10 V bias during irradiation and from the sensor without bias voltage is negligible, and thus it is not shown here.

\subsubsection{Interpixel capacitance}

The ``ring-connected'' mini-sensors were used to investigate the interpixel capacitance. The interpixel capacitance were determined by measuring the capacitance between the central pixel and the first surrounding pixel ring as function of bias voltage, while keeping the other two pixel rings and CCR grounded. Three frequencies of 10 kHz, 100 kHz and 1 MHz have been chosen for the interpixel capacitance measurement.

\begin{figure}[htbp]
\small
\centering
\includegraphics[width=7.5 cm]{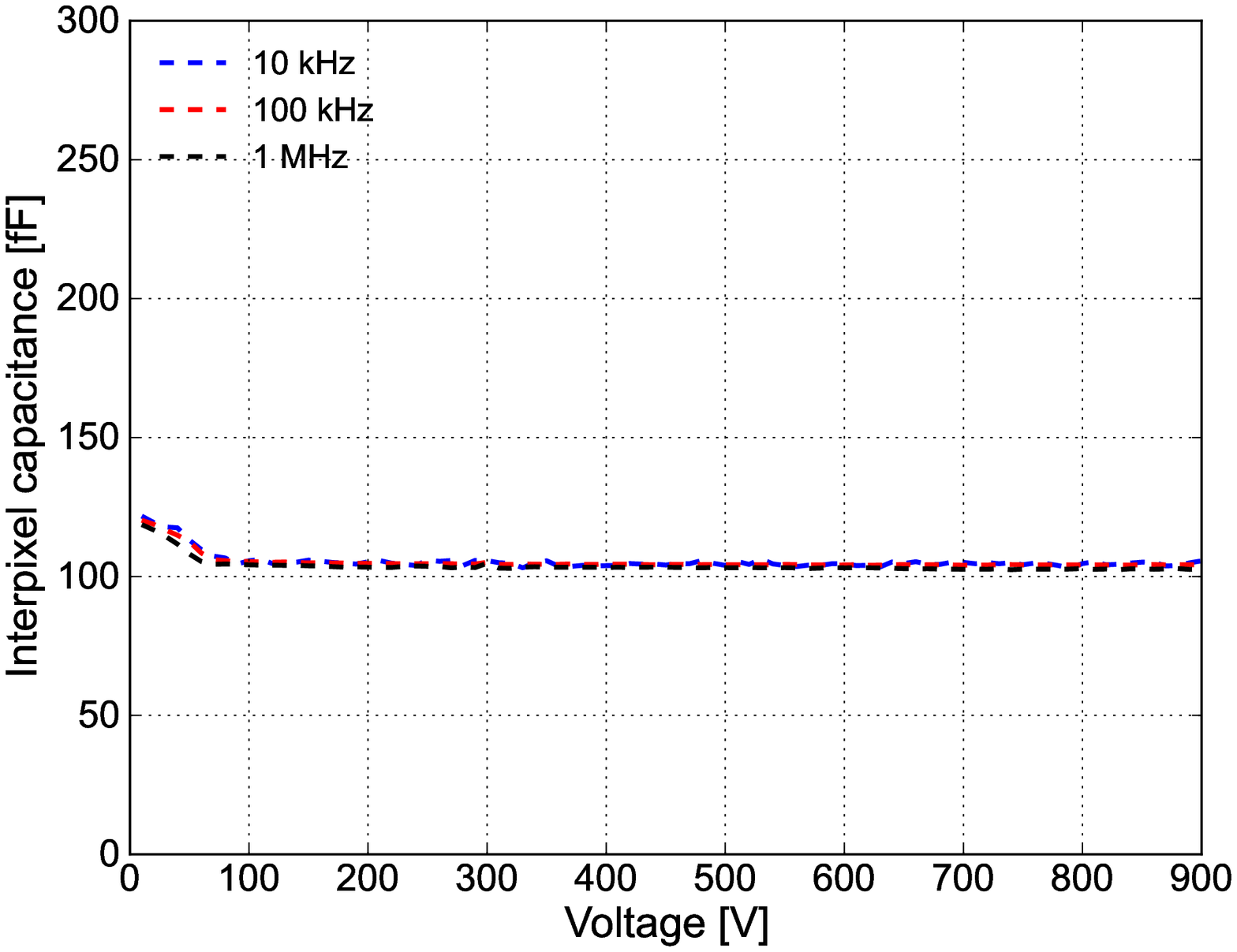}
\includegraphics[width=7.5 cm]{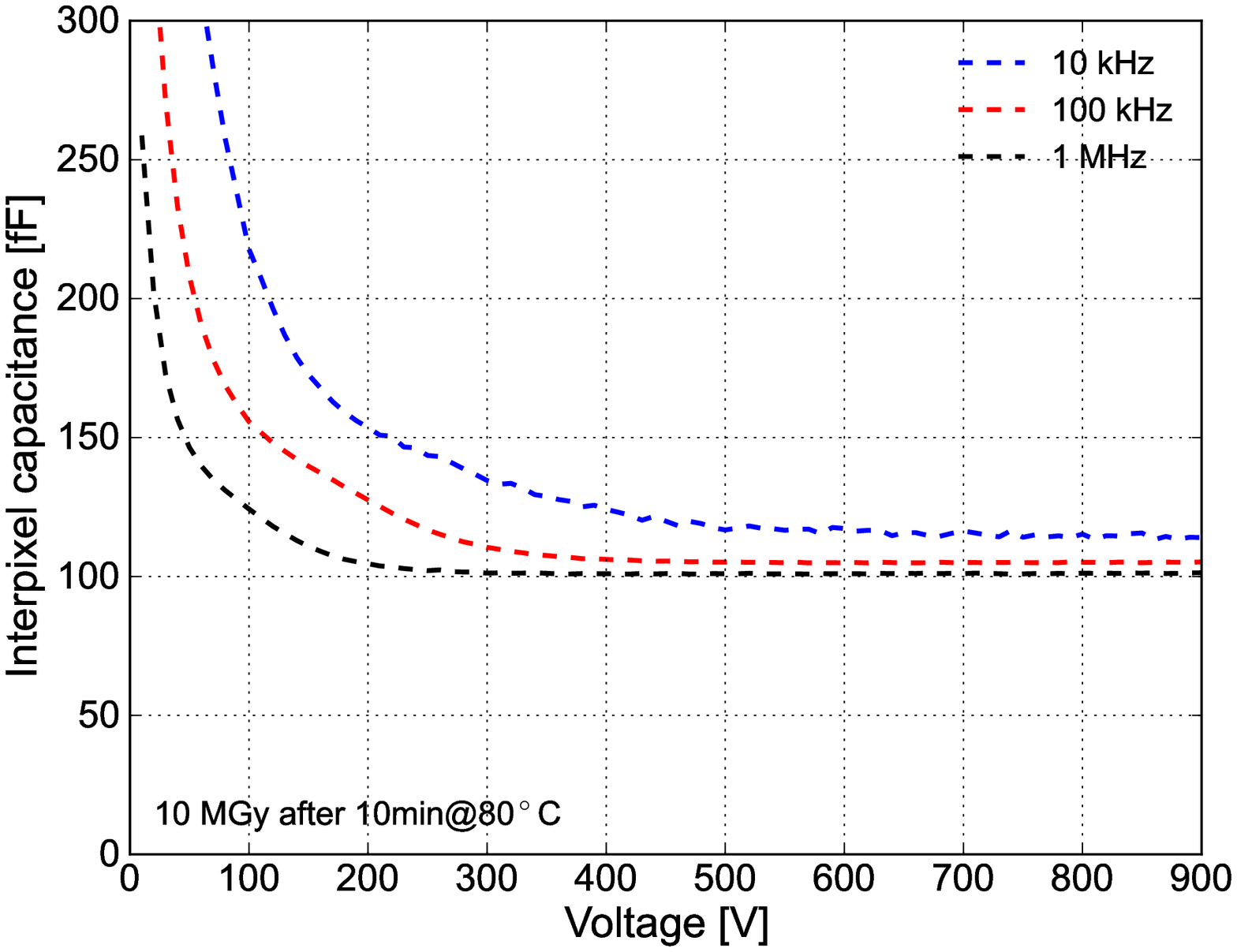}
\includegraphics[width=7.5 cm]{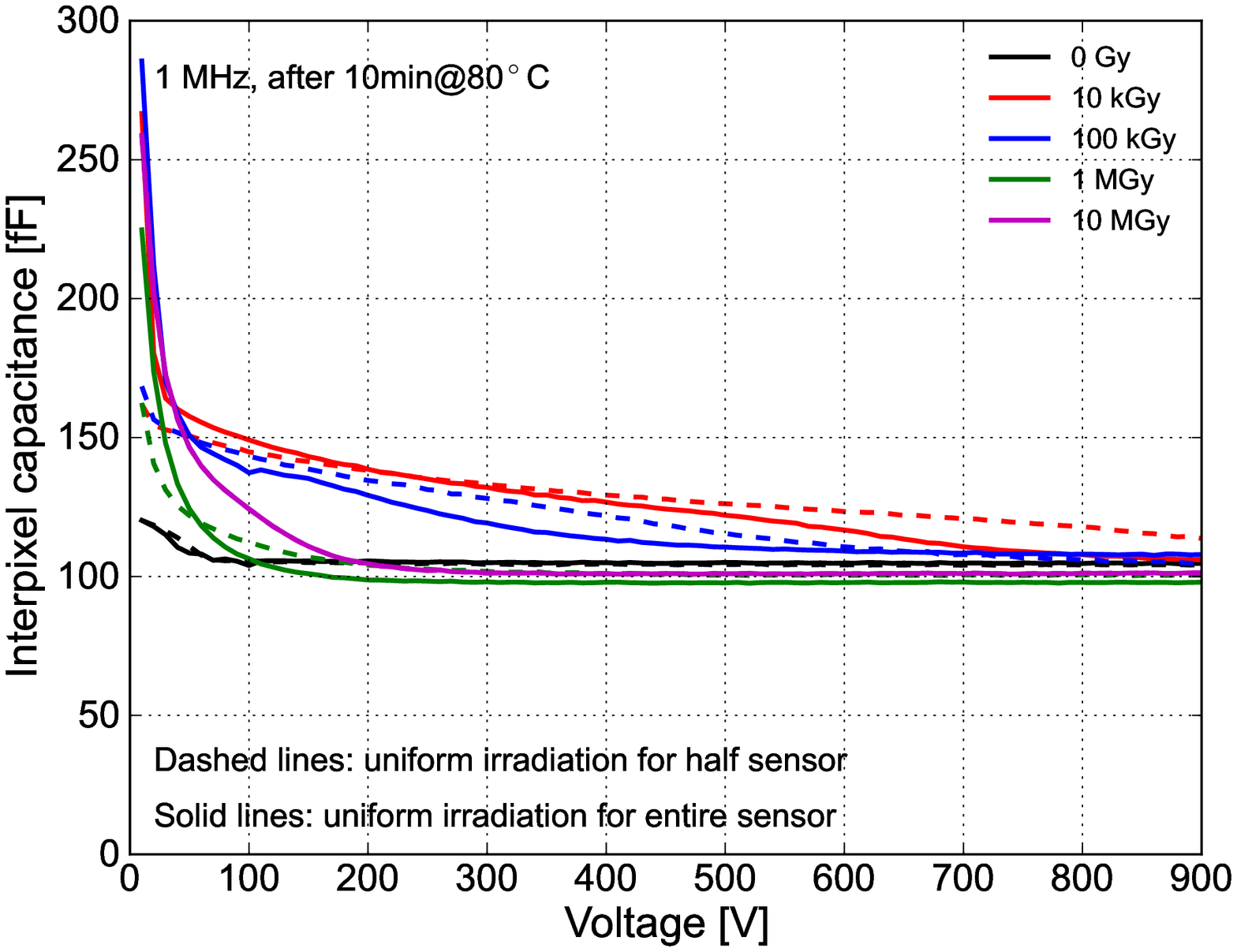}
\includegraphics[width=7.5 cm]{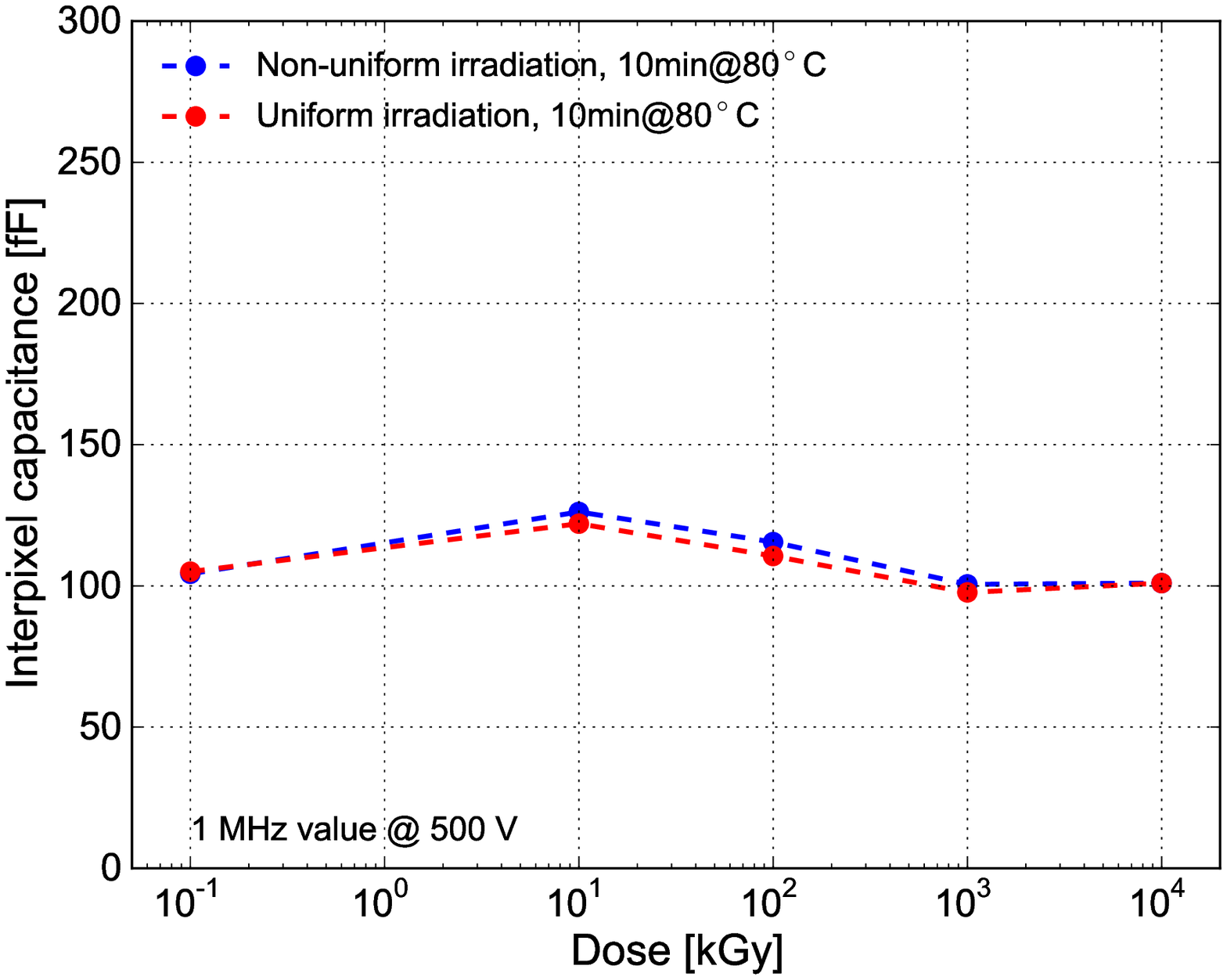}
\caption{Top left: $C_{int}$-$V$ before irradiation. Top right: $C_{int}$-$V$ after irradiation to 10 MGy and annealing at 80$^{\circ}$C for 10 minutes. Bottom left: $C_{int}$-$V$ as function of dose after annealing at 80$^{\circ}$C for 10 minutes. Bottom right: Dose dependence of interpixel capacitance at 500 V.}
\label{CintV}
\end{figure}

Before irradiation, as shown in figure \ref{CintV} (top left), there is no frequency dependence for the interpixel capacitance and the interpixel capacitance saturates at \mbox{$\sim$ 0.1 pF} at the full depletion voltage. However, a frequency dependence has been found after irradiation. As shown in figure \ref{CintV} (top right), the difference of the interpixel capacitance for different frequencies is 20\% to 30\% at 500 V, but the difference gets significant at lower bias voltage. The larger frequency spread indicates a larger contribution of the capacitance from the interface traps at lower bias voltage. As the sensor will be operated at a bias voltage of higher than 500 V, this will not be a concern for sensor operation.

Figure \ref{CintV} (bottom left) shows the interpixel capacitance vs. bias voltage at 1 MHz for different doses, measured from the ``ring-connected'' mini-sensors irradiated uniformly for the entire sensor and for half of the sensor. It is found that the interpixel capacitance decreases as function of bias voltage and saturates at a bias voltage higher than the full depletion voltage (some at > 900 V). All saturation values are $\sim$ 0.1 pF, which is similar to the value obtained before irradiation; the values at 500 V for different doses, as seen in figure \ref{CintV} (bottom right), are different by 30\% but still significantly smaller than the specification value of 0.5 pF.


\section{Summary and outlook}

The AGIPD sensor has been designed and fabricated by SINTEF. The first batch of AGIPD sensors has been received in the beginning of February of 2013. Radiation tests have been performed for the test structures and mini-sensors with the same pixel and guard-ring geometries as the AGIPD sensor. They have been irradiated to 10 kGy, 100 kGy, 1 MGy and 10 MGy with 8 keV and 12 keV monoenergetic X-rays at the beamline P11 of PETRA III. From the test structures, the saturation values of the oxide-charge density and the surface-current density for different irradiation doses are about \mbox{$(2-2.5) \times 10^{12}$ cm$^{-2}$} and \mbox{1.7 $\mu$A/cm$^{2}$}, which are similar to previous observations from the test structures produced by SINTEF with a thicker oxide layer. These values are reached at doses of \mbox{100 kGy} to \mbox{1 MGy}. From the mini-sensors, it is found that pixel current, CCR current, breakdown voltage and interpixel capacitance all are within specifications. The measurements demonstrate the radiation hardness of the sensor design for the AGIPD.

\acknowledgments

        The work was performed within the AGIPD project. I. Kopsalis would like to thank the German Academic Exchange Service "DAAD" for his PhD funding. We would like to thank the colleagues within the AGIPD collaboration for their helpful discussions on the results, and A. Meents and \mbox{A. Burkhardt} for the significant support for X-ray irradiations at the beamline P11 of DESY PETRA III. The work was also supported by the European XFEL company.


\begin{thebibliography}{21}

\bibitem{bib1}
M. Altarelli et al., \emph{XFEL: The European X-Ray Free-Electron Laser}, {\emph{Technical Design Report}, Preprint DESY 2006-097, DESY Hamburg 2006, ISBN 978-3-935702-17-1}, and \mbox{http://www.xfel.eu/de}.

\bibitem{bib2}
M.M. Seibert et al., \emph{Single mimivirus particles intercepted and imaged with an X-ray laser}, {\emph{Nature} {\bf 470} (2011) 78-82}, \mbox{doi: 10.1038/nature09748}.

\bibitem{bib3}
B. Henrich et al., \emph{The adaptive gain integrating pixel detector AGIPD - a detector for the European XFEL}, {\emph{Nucl.\ Instr.\ and\ Meth.\ A} {\bf 633} (2011) S11-S14}.

\bibitem{bib4}
D. Greiffenberg, \emph{The AGIPD detector for the European XFEL}, {\emph{JINST}  {\bf 7} (2012) C01103}.

\bibitem{bib5}
SYNOPSYS TCAD, http://www.synopsys.com/.

\bibitem{bib6}
J. Schwandt et al., \emph{Optimization of the radiation hardness of silicon pixel sensors for high x-ray doses using TCAD simulations}, {\emph{JINST}  {\bf 7} (2012) C01006}, \mbox{arXiv: 1111.4901}.

\bibitem{bib7}
J. Schwandt et al., \emph{Design of the AGIPD sensor for the European XFEL}, {\emph{JINST}  {\bf 8} (2013) C01015}, \mbox{arXiv: 1210.0430}.

\bibitem{bib8}
J. Zhang et al., \emph{Study of radiation damage induced by 12 keV X-rays in MOS structures built on high-resistivity n-type silicon}, {\emph{J.\ Synchrotron\ Rad.}  {\bf 19} (2012) 340-346}, \mbox{arXiv: 1107.5949}.

\bibitem{bib9}
J. Zhang et al., \emph{X-ray induced radiation damage in segmented p$^{+}$n silicon sensors}, {\emph{PoS Vertex2012} (2013) 019}.

\bibitem{bib10}
SINTEF ICT, http://www.sintef.no/.

\bibitem{bib11}
H. Graafsma, \emph{Requirements for the development of 2 dimensional X-ray detectors for the European X-ray Free Electron Laser in Hamburg}, {\emph{JINST}  {\bf 4} (2009) P12011}.

\bibitem{bib12}
J. Becker, \emph{Signal development in silicon sensors used for radiation detection}, {\emph{PhD thesis}, {\emph{University of Hamburg}}, {\bf DESY-THESIS-2010-033} (2010)}.

\bibitem{bib13}
J. Becker, \emph{Impact of plasma effects on the performance of silicon sensors at an X-ray FEL}, {\emph{Nucl. Instr. Meth.} A {\bf 615} (2010) 230-236, doi:10.1016/j.nima.2010.01.082}.

\bibitem{bib14}
J. Becker, \emph{Simulation and experimental study of plasma effects in planar silicon sensors}, {\emph{Nucl. Instr. Meth.} A {\bf 624} (2010) 716-727, doi:10.1016/j.nima.2010.10.010}.

\bibitem{bib15}
E.H. Nicollian, \emph{MOS (Metal Oxide Semiconductor Physics and Technology)}, {\emph{New York: John Wiley and Sons} (1982) ISSN 0-471-08500-6}.

\bibitem{bib16}
D.M. Fleetwood, \emph{Defects in Microelectronic Materials and Devices}, {\emph{Boca Raton: Taylor and Francis} (2008) ISBN 978-1-4200-4376-1}.

\bibitem{bib17}
J. Zhang et al., \emph{Investigation of X-ray induced radiation damage at the Si-SiO$_{2}$ interface of silicon sensors for the European XFEL}, {\emph{JINST}  {\bf 7} (2012) C12012}, \mbox{arXiv: 1210.0427}.

\bibitem{bib18}
J. Zhang, \emph{X-ray Radiation Damage Studies and Design of a Silicon Pixel Sensor for Science at the XFEL}, {\emph{PhD thesis}, {\emph{University of Hamburg}}, {\bf DESY-THESIS-2013-018} (2013)}.

\bibitem{bib19}
H. Perrey, \emph{Jets at Low Q$^{2}$ at HERA and Radiation Damage Studies for Silicon Sensors for the XFEL}, {\emph{PhD thesis}, {\emph{University of Hamburg}}, {\bf DESY-THESIS-2011-021} (2011)}.

\bibitem{bib20}
The webpage of pco.: http://www.pco.de.

\bibitem{bib21}
J. Zhang et al., \emph{Study of X-ray Radiation Damage in Silicon Sensors}, {\emph{JINST}  {\bf 6} (2011) C11013}, \mbox{arXiv: 1111.1180}.


\end{thebibliography}
\end{document}